\documentclass[runningheads]{llncs}

\usepackage[utf8]{inputenc}
\usepackage{caption}
\usepackage{xspace}
\usepackage{hyperref}
\usepackage{wrapfig}
\usepackage{floatflt}
\usepackage{listings}
\usepackage[ruled]{algorithm2e}
\usepackage{tikz}
\usepackage{amsfonts}
\usepackage{amsmath}
\usepackage{amssymb}
\usepackage{mathtools}
\usepackage{float}
\usepackage{subcaption}
\usepackage{lineno}
\usepackage{calc}
\usepackage{etoolbox}
\usepackage{array}
\usepackage{color, colortbl}
\usepackage{multirow}
\usepackage{arydshln}
\usepackage{enumitem}
\usepackage[style=lncs]{biblatex} 
\usepackage{breqn}

\floatstyle{boxed}
\newfloat{boxed_table}{h}{modal_props.aux}
\floatname{boxed_table}{Table}

\providecommand{\keywords}[1]
{
  \small	
  \textbf{\textit{Keywords---}} #1
}
\newcounter{theoremcnt}[section]
\renewcommand{\thetheoremcnt}{\arabic{theoremcnt}}

\newenvironment{invariant}
{\begin{trivlist}\refstepcounter{theoremcnt}
\item[]\bf Invariant~\thetheoremcnt.~\rm}{\end{trivlist}}

\newcommand{\tab}[2]{\setlength{\hspace*}{0.4cm * #2}#1\setlength{\hspace*}{0.4cm - \widthof{#1}}}
\newcommand{\true}{\textit{true}}
\newcommand{\false}{\textit{false}}
\newcommand{\Nat}{\mathbb{N}}

\newcommand{\sump}{\sum_{p{:}P}}
\newcommand{\etc}{\textit{etc}.\rangle}
\newcommand{\inta}{\textit{int}\xspace}

\title{Verification of the busy-forbidden protocol\thanks{This publication is part of the MasCot PVSR project (00795160) which is partly financed by the Dutch Research Council (NWO).} \vspace{0.3ex}\\
\small(using an extension of the cones and foci proof framework)}
\author{P.H.M. van Spaendonck \orcidID{0000-0002-9536-1524}}

\institute{Department of Mathematics and Computer Science,\\
    Eindhoven University of Technology\\
    \texttt{P.H.M.v.Spaendonck@tue.nl}}

\date{}
\titlerunning{Verification of the busy-forbidden protocol}

\begin{document}
\maketitle
\begin{abstract}
The busy-forbidden protocol is a new readers-writer lock with no resource contention between readers, which allows it to outperform other locks. 
For its verification, specifications of its implementation and its less complex external behavior are provided by the original authors but are only proven equivalent for up to $7$ threads.

We provide a general equivalence proof using the cones and foci proof framework, which rephrases whether two specifications are branching bisimilar as six properties on the data objects of the specifications.
We provide an extension of this framework consisting of four additional properties and prove that when the additional properties hold, the two systems are divergence-preserving branching bisimilar, a stronger version of the aforementioned relation that also distinguishes livelocks.

\end{abstract}

\section{Introduction}
The readers-writer lock problem is a concurrency problem introduced and solved by Courtois et al.\ \cite{rwlock}.
The problem requires a synchronisation protocol that provides safe access to both a shared readers section, which can be used simultaneously by any number of threads, as well as an exclusive writer section, which can not be used by more than one thread at any given time and only when the readers section is not in use.

In \cite{paraterm}, Groote et al.\ introduce a new readers-writer lock called the busy-forbidden protocol. 
This locking protocol is of particular interest as it has no resource contention between readers, and therefore provides a significant speedup over other locks when having high readers section workloads.

To ensure the correctness of the protocol, the authors give process algebraic specifications of both the implementation of the new algorithm as well as a specification of its external behavior.
The authors applied model checking and proved the implementation and external behavior equivalent for up to $7$ threads using the mCRL2 toolset \cite{mcrl2toolset}, but they were unable to do this for more concurrent threads due to the statespace of the implementation becoming too large.

But as readers-writer locks often use a large number of concurrent threads, a general correctness proof for the busy-forbidden protocol is desired.
We opt to prove the process algebraic specifications of the implementation and external behavior to be equivalent.
The advantage of this technique over contract-based approaches, such as Floyd-Hoare logic \cite{hoare}, and its extension for parallel composed systems by Owicki and Gries \cite{owicki1976axiomatic, owicki1976verifying}, is that the much smaller equivalent model can also be used for the modeling and verification of systems built on top of the busy-forbidden protocol.
We consider this a significant advantage, as this is the typical use-case for readers-writer locks, e.g.\ the parallel term library which the protocol was originally designed for.

We prove the equivalence of the implementation and its external behavior by using the cones and foci proof framework, originally proposed in \cite{Grootespringintveld} by Groote and Springintveld and later generalized by Fokkink et al.\ in \cite{Fokkink}.
This framework simplifies the often complex and cumbersome branching bisimulation proof by reducing it to a small set of propositions on the data objects occurring in the implementation and specification. If these propositions are shown to hold, it follows that the two systems are equivalent modulo branching bisimulation.

The proof framework has already been used in several case studies to prove implementation and specification models equivalent, such as the verification of the 1-bit sliding window protocol in \cite{slidingwindow}, a complex leader election protocol in \cite{FREDLUND1997459}, and a part of the IEEE P1394 high-speed bus protocol \cite{P1394} in \cite{Shankland}.

Since the equivalence relation proven by the cones and foci proof framework does not distinguish livelock, we first provide an extension to the framework such that it can also be used to prove equivalence modulo divergence-preserving branching bisimulation.
This relation is a stronger version of branching bisimulation that does distinguish livelocks \cite{lmcs:758}.
Our extension provides four additional propositions on the data objects in the implementation and specification models, that, when shown to also hold, imply the equivalence of the two processes modulo divergence-preserving branching bisimulation.
We give a soundness proof of this extension and use it to prove the equivalence of implementation and specification of the busy-forbidden protocol.

\section{The busy-forbidden protocol} \label{sec:model} 
We first discuss the busy-forbidden protocol.
An overview of its implementation using pseudocode is given in Table \ref{table:busy_forbidden}.
The \texttt{enter\_-} and \texttt{leave\_shared} functions are used to have a thread $p$ enter or leave the readers section.
Similarly, \texttt{enter\_-} and \texttt{leave\_exclusive} provide functionality for safe access to the writer section.

The protocol uses two binary flags per thread and a single mutex. The first flag, the \textit{busy} flag, indicates that a thread is either working or going to work inside of the readers section.
The second flag, the \textit{forbidden} flag, indicates that a thread is not allowed to enter the readers section.
All flags are initially set to $\false$.
The mutex, called \textit{mutex}, enforces exclusive access to the writer section.

\begin{table}[h]
\begin{center}
\begin{tabular}{|l|l|}
\hline
\begin{minipage}{0.4\textwidth}
$\texttt{enter\_shared}(\textit{thread}~p):$\\
\hspace*{0.4cm}$p.\textit{busy}:=\true$;\\
\hspace*{0.4cm}$\textbf{while}~p.\textit{forbidden}$\\
\hspace*{0.8cm}$p.\textit{busy}:=\false;$\\
\hspace*{0.8cm}$\textbf{if}~\textit{mutex}.\textit{timed\_lock}()$\\
\hspace*{1.2cm}$\textit{mutex}.\textit{unlock}();$\\
\hspace*{0.8cm}$p.\textit{busy}:=\true;$\\
\\
\end{minipage}&
\begin{minipage}{0.4\textwidth}
$\texttt{enter\_exclusive}(\textit{thread}~p):$\\
\hspace*{0.4cm}$\textit{mutex}.\textit{lock}();$\\
\hspace*{0.4cm}$\textbf{while exists}~\textit{thread}~q\textbf{ with}$\\
\hspace*{1.2cm}$\neg q.\textit{forbidden}$\\
\hspace*{0.8cm}$    \textbf{select}~\textit{thread}~r$\\
\hspace*{0.8cm}$      r.\textit{forbidden}:=\true;$\\
\hspace*{0.8cm}$    \textbf{if}~r.\textit{busy} ~\textbf{or sometimes}$\\
\hspace*{1.2cm}$        r.\textit{forbidden}:=\false;$\\
\end{minipage}\\
\hline
\begin{minipage}{0.4\textwidth}
$\texttt{leave\_shared}(\textit{thread}~p):$\\
\hspace*{0.4cm}$p.\textit{busy}:=\false;$\\
\\\\\\\\
\\\\\\
\end{minipage}&
\begin{minipage}{0.4\textwidth}
$\texttt{leave\_exclusive}(\textit{thread}~p):$\\
\hspace*{0.4cm}$\textbf{while exists}\textit{ thread q}\textbf{ with}$\\
\hspace*{1.2cm}$q.\textit{forbidden}$\\
\hspace*{0.8cm}$\textbf{select }\textit{thread } r$\\
\hspace*{0.8cm}\textbf{usually do}\\
\hspace*{1.2cm}$ r.\textit{forbidden}:=\false;$\\
\hspace*{0.8cm}\textbf{sometimes do}\\
\hspace*{1.2cm}$ r.\textit{forbidden}:=\true$\\
\hspace*{0.4cm}$mutex.unlock();$\\
\end{minipage}\\
\hline
\end{tabular}
\end{center}
\caption{Pseudocode description of the busy-forbidden protocol.}
\label{table:busy_forbidden}
\end{table}

When entering the readers section, a thread sets its \textit{busy} flag and enters iff its \textit{forbidden} flag is $\false$.
If the \textit{forbidden} flag is $\true$, the \textit{busy} flag is set back to $\false$ to avoid deadlock and the process is repeated again.
To reduce resource contention on the flags, a $\textit{mutex}.\textit{timed\_lock}()$ can be used without altering the externally visible behavior of the protocol \cite{paraterm}.
Upon leaving the readers section, the thread sets its \textit{busy} flag back to false.

A thread that wants to enter the writer section must first acquire the mutex.
This ensures that no other thread can be in the writer section simultaneously and that only the given thread is altering the \textit{forbidden} flags.
Once the mutex has been acquired, the thread sets the \textit{forbidden} flag of each thread, but will immediately undo this if the \textit{busy} flag of the same thread is $\true$.
To prevent a thread that is acquiring the writer section from locking out some reader threads while still waiting for others to leave the readers section, random \textit{forbidden} flags can sometimes be set back to $\false$.
The writer section is entered once all \textit{forbidden} flags are $\true$.
Upon leaving, all \textit{forbidden} flags are set back to false and the mutex is released.
During this, random \textit{forbidden} flags can be set back to $\true$.
This prevents each iteration that occurs while leaving, from becoming externally visible and significantly reduces the number of states in the external specification.

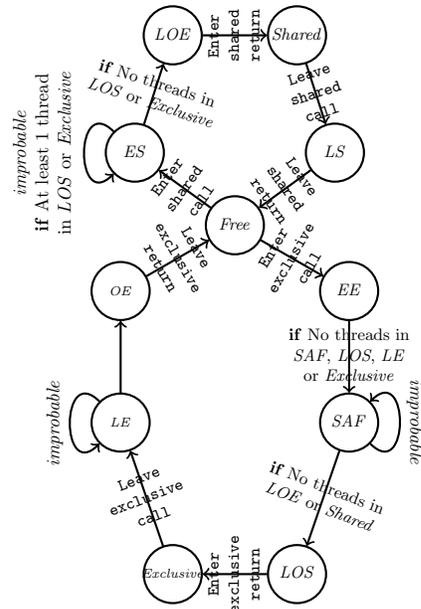
\begin{wrapfigure}{r}{0.5\textwidth}
\centering
\newcommand{\state}[4]{
     \node[thick, circle, minimum size=0.1cm, draw] (#3) at (#1,#2) 
     {\raisebox{0cm}[1ex][0ex]{\makebox[0.8cm]{\small\textit{#4}}}};}
\newcommand{\trans}[4]{\draw[thick, ->] (#1) -- (#2) node[midway, rotate=#4]{\small\begin{tabular}{c}#3\end{tabular}};}
\newcommand{\tauloop}[2]{
    \draw[->, thick] (#1) 
        edge[in=\ifthenelse{\ifstrequal{#2}{left}}{45}{225}, out=\ifthenelse{\ifstrequal{left}{#2}}{-45}{-224}]
        node[#2]
        (#1);
}
\begin{tikzpicture}[scale=0.7, every node/.style={transform shape}]
  \state{ 0.00}{0.00}{Free}{Free};
  \state{-1.90}{1.38}{EnteringShared}{ES}
  \state{-1.18}{3.61}{LockedOutExclusive1}{LOE}
  \state{ 1.18}{3.61}{Shared}{Shared}
  \state{ 1.90}{1.38}{LeavingShared}{LS}
  \trans{Free}{EnteringShared}{\texttt{Enter}\\\texttt{shared}\\\texttt{call}}{54}
  \trans{EnteringShared}{LockedOutExclusive1}{\textbf{if} No threads in\\ \textit{LOS} or \textit{Exclusive}}{-18}
  \draw[->, thick] (EnteringShared)
    edge[in=225, out=-225, looseness=3.4]
    node[above, rotate = 90]
    {\begin{tabular}{c}
        \textit{improbable} \\
        \textbf{if} At least 1 thread\\
        in \textit{LOS} or \textit{Exclusive}
    \end{tabular}}
    (EnteringShared);
  \trans{LockedOutExclusive1}{Shared}{\texttt{Enter}\\\texttt{shared}\\\texttt{return}}{90}
  \trans{Shared}{LeavingShared}{\texttt{Leave}\\\texttt{shared}\\\texttt{call}}{18}
  \trans{LeavingShared}{Free}{\texttt{Leave}\\\texttt{shared}\\\texttt{return}}{-54}
  \state{+2.17}{-1.25}{EnteringExclusive}{EE}
  \state{+2.17}{-3.75}{LockedOutExclusive2}{SAF}
  \state{+1.18}{-6.63}{LockedOutAll}{LOS}
  \state{-1.18}{-6.63}{Exclusive}{\scriptsize Exclusive}
  \state{-2.17}{-3.75}{LeavingExclusive1}{\scriptsize LE}
  \state{-2.17}{-1.25}{LeavingExclusive2}{\scriptsize OE}
  \trans{Free}{EnteringExclusive}{\texttt{Enter}\\\texttt{exclusive}\\\texttt{call}}{60}  
  \trans{EnteringExclusive}{LockedOutExclusive2}{\textbf{if} No threads in\\\textit{SAF}, \textit{LOS}, \textit{LE}\\or \textit{Exclusive}}{0}  
  \draw[thick, ->] (LockedOutExclusive2) edge[in=45, out=-45, looseness=3.4] node[above, rotate = -90] {\textit{improbable}} (LockedOutExclusive2);
  \trans{LockedOutExclusive2}{LockedOutAll}{\textbf{if} No threads in\\\textit{LOE} or \textit{Shared}}{-21}  
  \trans{LockedOutAll}{Exclusive}{\texttt{Enter}\\\texttt{exclusive}\\\texttt{return}}{90}
  \trans{Exclusive}{LeavingExclusive1}{\texttt{Leave}\\\texttt{exclusive}\\\texttt{call}}{21}
  \trans{LeavingExclusive1}{LeavingExclusive2}{}{0}
  \draw[thick, ->] (LeavingExclusive1) edge[in=225, out=-225, looseness=3.4] node[above, rotate = 90] {\textit{improbable}}(LeavingExclusive1);
  \trans{LeavingExclusive2}{Free}{\texttt{Leave}\\\texttt{exclusive}\\\texttt{return}}{-60}  
\end{tikzpicture}
\caption{The external behaviour.}
\label{fig:external-busy-forbidden}
\end{wrapfigure}

The externally visible behavior of the protocol is given in Figure \ref{fig:external-busy-forbidden} and, as we will prove later, provides an equivalent overview of how threads interact via the protocol.
Individual threads move from node to node.
Transition labels ending with \texttt{-call} represent the identically named function being called by a thread moving across, and those ending with \texttt{-return} represent those function calls terminating.
All transitions not labeled as such represent some sequence of internal calculations that occurs during these function calls.
Transitions labeled with a guard, i.e.\ starting with \textbf{if}, only allow a thread to progress if the given condition is met.

The \textit{Free} node represents a thread not interacting with the protocol and being outside of any section.
Each thread initially starts out in this node.
The \textit{Shared} and \textit{Exclusive} nodes represent the readers and writer sections, respectively.

A thread starting to acquire the readers lock enters the \textit{EnterShared} (\textit{ES}) node.
The thread stays in the \textit{ES} node as long as its $\textit{forbidden}$ flag is $\true$.
As repeatedly checking the flag is discouraged through the \textit{timed\_lock} call, the internal loop is labeled as \textit{improbable}.
When the $\textit{forbidden}$ flag is evaluated to $\false$, the thread moves to the \textit{LockedOffExclusive} (\textit{LOE}) node. 
After this, it is no longer possible for any other thread to enter the writer section until the readers section is completely freed.
The \textit{LeaveShared} (\textit{LS}) node represents a thread leaving this section.

When a thread tries to acquire the writer lock, it enters the \textit{EnterExclusive} (\textit{EE}) node.
Once the thread acquires the \textit{mutex} variable, it will move to the \textit{SetAllForbidden} (\textit{SAF}) node and it will not be possible for any other thread to acquire the writer lock before it is released by this thread.
The loop in the \textit{SAF} node represents a \textit{forbidden} flag being set back to $\false$; this transition is labeled as \textit{improbable} as this only rarely occurs.
Once the last $\textit{busy}$ flag is evaluated to $\false$, exclusive access is attained and the thread will move to the \textit{LockedOutShared} (\textit{LOS}) node before officially terminating the function call.

When the thread starts releasing the writer lock, it enters the \textit{LeavingExclusive} (\textit{LE}) node.
Similar to the \textit{SAF} node, a thread within the \textit{LE} node can repeatedly turn the \textit{forbidden} flag off and on again, thus never fully opening up the readers section.
Because a \textit{forbidden} flag is only very rarely set back to $\true$ when releasing the lock, this transition is also labeled as \textit{improbable}.
Once the last \textit{forbidden} flag is set to $\false$, this is no longer possible and the thread moves to the \textit{OpenedExclusive} (\textit{OE}) node, after which it will officially terminate the function call and move back to the \textit{Free} node.

We can use the model of the external behavior to reason about certain safety properties.
For example, from the guarded transitions from \textit{ES} to \textit{LOE} and from \textit{SAF} to \textit{LOS}, we can quickly see that the \textit{Shared} and \textit{Exclusive} sections can not be populated simultaneously, as they require the other respective section to be empty.
The guarded transition from \textit{EE} to \textit{SAF} also assures that only a single thread can be present in the \textit{Exclusive} section at any given time.


\section{Linear process equations}
Both the implementation of the pseudocode shown in Table \ref{table:busy_forbidden} and the external behavior have been modeled in the mCRL2 language \cite{mcrl2}. The mCRL2 language is based on the Algebra of Communicating Processes \cite{DBLP:books/daglib/0069083} and Calculus of Communicating Processes \cite{DBLP:books/sp/Milner80}.

The mCRL2 language models processes using a combination of states and actions.
States represent a collection of internal values that are used to calculate which actions can occur and what the resulting state will be.
Actions represent any sort of atomic event such as calling a function, or setting or reading a flag.
An action consists of a label and a possible set of data parameters, e.g.\ the action $\textit{lock}(p)$ has $\textit{lock}$ as the label and $p$ as the data parameter.
Parameters can be of varying types such as booleans, algebraic data types, and mappings.
The exact data types used within the busy-forbidden models are given later.

A special action $\tau$, the so-called hidden or internal action, is used to represent an action that is externally not directly visible.
We use distinct action labels for internal actions to be able to easily distinguish between them.
We explicitly state which actions should be considered to be $\tau$ actions.

We require all process algebraic equations to be in a clustered linear form, see Definition \ref{def:lps}.
This form specifies for each action when it can occur and what the resulting state will be.
The $\sum_{e{:}S}$ operator models the application of the non-deterministic choice operator $+$ over all elements in some set $S$.
We also allow process equations in which the $\sum$ operators are split into separate smaller $\sum$ operators and individual $+$ operators. 

Since the cones and foci proof framework concerns itself only with the actions that are enabled in a single given state, the clustered normal form becomes especially useful, as we can directly infer for any given state if an action is enabled and what the resulting state will be.
In \cite{Usenko}, Usenko shows that any mCRL2 specification can be transformed into a clustered linear process equation.

\begin{definition} \label{def:lps}
    \normalfont{A clustered linear process equation (LPE) is a process specification of the form:}
    $$X(d{:}D) = \sum_{a{:}\textit{Act}} \, \sum_{e_a{:}E_a} c_a(d,e_a) \to a(f_a(d,e_a)) \cdot X(g_a(d,e_a)) \text{,}$$
    \normalfont{where $D$ is the set of states, $\textit{Act}$ is the set of action labels including $\tau$, $E_a$ is an indexed set of all data types that need to be considered for label $a$,  the boolean function $c_a(d, e_a)$ specifies when the action $a$ with parameters resulting from the function $f_a(d,e_a)$ is enabled in state $d$, and $g_a(d, e_a)$ gives the resulting state from taking this action from state $d$.}
\end{definition}

Often we end up in a situation in which the set of states $D$ also contains unreachable states.
As we are only interested in the reachable states, we introduce the notion of an invariant in Definition \ref{def:invariant}.
An invariant is a predicate on states in an LPE such that when it holds for a given state $d{:}D$, it also holds for all subsequent states.
\begin{definition} \label{def:invariant}
    \normalfont{Given a clustered \textit{LPE} $X$ as per Definition \ref{def:lps}.
    A predicate $\mathcal{I}$ on the set of states $D$ is called an invariant iff the following holds: for all $a{:}\textit{Act}, d{:}D$ and $e_a{:}E_a$,}
    $$\mathcal{I}(d) \wedge c_a(d, e_a) \Rightarrow \mathcal{I}(g_a(d, e_a))$$
\end{definition}

\section{Equivalence and the cones and foci proof framework}
As stated before, we prove the model of the implementation and the specification of the busy-forbidden protocol equivalent modulo divergence-preserving branching bisimulation. 
We define this equivalence relation in Definition \ref{def:dpbb}, which is based on the definitions used in \cite{luttik} and has been adapted to work with process equations instead of transition systems. In Definition \ref{def:lpe2lts}, we provide some syntactic glue to make this shift between labeled transition systems and clustered LPEs more intuitive.

\begin{definition} \label{def:lpe2lts}
    \normalfont{Given a clustered LPE as per Definition \ref{def:lps}, states $d,d' \in D$, and action \textit{l}, we define the following relations:
    \begin{itemize}
        \item $d \xrightarrow{l} d'$ iff there is an action $a$ with an associated data type $e_a$ such that $l = a(f_a(d,e_a))$, the condition $c_a(d,e_a)$ holds, and $g_a(d,e_a) = d'$.
        \item $d \xrightarrow{l}\!\!\!^* d'$ iff there is a finite sequence of states $d_0, \hdots, d_k$ such that $d_0 = d$, $d_k = d'$ and for all $0 \leq i < k$ we have $d_i \xrightarrow{l} d_{i+1}$.
        
    \end{itemize}}
\end{definition}

\begin{definition} \label{def:dpbb}
\normalfont{Given two clustered LPEs as per Definition \ref{def:lps} with sets of states $D$ and $D'$. A relation $R$ on the states $D \times D'$ is a divergence-preserving branching bisimulation iff the following conditions for all states $s \in D$, $t\in D'$, and actions $\textit{l} \in \textit{Act}$ hold:
\begin{itemize}
    \item[($\textit{B}_1$)] If $s R t$ and $s \xrightarrow{l} s'$ for some state $s'\in D$, then either $l = \tau$ and $s' R t$, or there are states $t', t'' \in D'$ such that $t \xrightarrow{\tau}\!^* t' \xrightarrow{l} t''$, $s R t'$, and $s' R t''$.
    \item[($\textit{B}_2$)] If $s R t$ and $t \xrightarrow{l} t'$ for some state $t'\in D'$, then either $l = \tau$ and $s R t'$, or there are states $s', s'' \in D$ such that $s \xrightarrow{\tau}\!^* s' \xrightarrow{l} s''$, $s' R t$, and $s'' R t$.
    \item[($\textit{D}_1$)] If $s R\, t$ and there is an infinite sequence of states $(s_n)_{n\in\mathbb{N}}$ such that $s = s_0$, and $s_k \overset{\tau}{\to}s_{k+1}$ and $s_k R\, t$ for all $k \in \mathbb{N}$, then there is a state $t' \in D'$ such that $t \overset{\tau}{\to} t'$, and $s_k R\, t'$ for some $k\in\mathbb{N}$.
    \item[($\textit{D}_2$)] If $s R\, t$ and there is an infinite sequence of states $(t_n)_{n\in\mathbb{N}}$ such that $t = t_0$, and $t_k \overset{\tau}{\to}t_{k+1}$ and $s R\, t_k$ for all $k \in \mathbb{N}$, then there is a state $s' \in D$ such that $s \overset{\tau}{\to} s'$, and $s' R\, t_k$ for some $k\in\mathbb{N}$.
\end{itemize}

\noindent Two clustered LPEs with respective initial states $d_0$ and $d_0'$ are \textit{divergence-preserving branching bisimilar} iff there is a divergence-preserving branching bisimulation $R$ such that $d_0 R\, d_0'$.}
\end{definition}

Note that in (divergence-preserving) branching bisimulation, $\tau$-actions are said to be externally visible iff their begin- and endpoint are not equivalent.

\begin{wrapfigure}{r}{0.45\textwidth}
    \centering
    \begin{tikzpicture}
    \draw[draw=none, fill=gray!20] (-5,1.5) -- (0,0) -- (-5,-1.5) -- cycle {};
    \draw[thick] (-5,1.5) -- (0,0) -- (-5,-1.5) {};
    \node (FC) at (0,0) {};
    \filldraw[black] (0,0) circle (2pt) node[anchor=west]{\textit{fc}};
    \filldraw[black] (-4,0.4) circle (2pt) node[anchor=east]{$d$};
    \draw[dashed, ->] (-4,0.4) -- (-3.5,1.05) {};
    \draw[->] (-3.5, 1.05) -- (-3.25, 1.45) node[midway, left] {$a$};
    \draw[dashed, ->] (-3.5,1.05) -- (-2.5,-0) {};
    \draw[dashed, ->] (-2.5,0) -- (-2, -0.6) {};
    \draw[->] (-2,-0.6) -- (-1.75, -1.0) node[midway, left] {$b$};
    \draw[dashed, ->] (-2, -0.6) -- (-1, 0) {};
    \draw[dashed, ->] (-1, 0) -- (0,0) {};
    
    \draw[->] (0,0) -- (0.25, 0.4) node[midway, left] {$a$};
    \draw[->] (-1,0.3) -- (-0.75, 0.7) node[midway, left] {$a$};
    \draw[->] (-2,0.6) -- (-1.75, 1.0) node[midway, left] {$a$};
    \draw[->] (-4.5,1.35) -- (-4.25, 1.75) node[midway, left] {$a$};
    
    \draw[->] (0,0) -- (0.25, -0.4) node[midway, left] {$b$};
    \draw[->] (-1,-0.3) -- (-0.75, -0.7) node[midway, left] {$b$};
    \draw[->] (-3.5,-1.05) -- (-3.25, -1.45) node[midway, left] {$b$};
    \draw[->] (-4.5,-1.35) -- (-4.25, -1.75) node[midway, left] {$b$};
    
    \node (conename) at (-4.5,-0.625) {\large $C$};
    \end{tikzpicture}
    \caption{A cone $C$ with focus point \textit{fc}.}
    \label{fig:cone}
\end{wrapfigure}
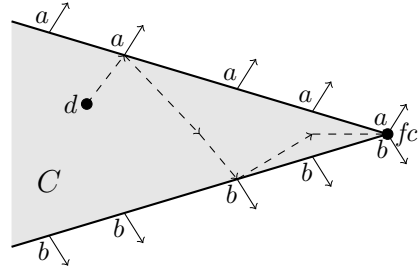

In \cite{Grootespringintveld}, it is noted that in communicating systems, equivalent states often have a 
``cone-like" structure as is shown in Figure \ref{fig:cone}.
In this figure, equivalent states are grouped together in the \textit{cone} $C$. 
In the \textit{focus point} state $\textit{fc}$, all externally visible actions of said cone, i.e. $a$ and $b$, are enabled. 
For all other states in which not all externally visible actions are simultaneously enabled, such as $d$ or the states along the edges, there is always a path of \textit{internal actions}, i.e. $\tau$ actions within the cone, that ends in the state $\textit{fc}$. We show one such path for the state $d$, using the dashed arrows.

If a given system consists of such ``cones'', the cones and foci proof framework can be used to prove equivalence. 
To do so, we must provide a \textit{state mapping} $h : D \to D'$ that maps states in the implementation to their equivalent state in the specification,
a \textit{focus condition} $\textit{FC} : D \to \mathbb{B}$ that indicates if a state should be considered a focus point, i.e.\ all externally observable actions are enabled,
and a well-founded ordering $<_M$ on $D$ that orders states by their distance to a focus point.
We must then prove that a small set of requirements are met by the LPEs and the provided \textit{state mapping, focus condition} and ordering.

Any $\tau$ action in the implementation that does not leave a cone, i.e.\ the state mapping $h$ maps begin- and endpoint to the same state, is renamed to \inta (short for \textit{internal} action). 
This allows us to easily distinguish between $\tau$ actions that are externally observable, i.e.\ that are preserved in our specification, and those that are not. 
While an \inta action is considered a $\tau$ action, we exclude them from the set of actions \textit{Act}.

In Theorem \ref{thm:new_cnf}, we extend the proof framework towards divergence-preserving branching bisimulation with a labeling $p$ on cones that labels cones as either divergent $(\Delta)$ or non-divergent $(\nabla)$, and four additional requirements on the LPEs.
The divergent $\tau$-loops in the specification, i.e.\ a $\tau$ transition with the same begin- and endpoint, are renamed to \inta to relate these to the divergent internal behavior in the implementation, i.e.\ repeatable paths of $\inta$ actions.

\begin{theorem} \label{thm:new_cnf}
\normalfont{Consider a clustered linear process equation of an implementation with initial state $d_0$ and some invariant $\mathcal{I}$ that holds in $d_0$,}

$$X(d{:}D) = \sum_{a{:} \textit{Act}\cup\{\inta\}} \sum_{e_a{:} E_a} c_a(d,e_a) \rightarrow a(f_a(d, e_a)) \cdot X(g_a(d, e_a))\text{,}$$

\noindent \normalfont{and a clustered linear process equation of a specification with initial state $d_0'$,}

$$Y(d'{:}D') = \sum_{a{:} \textit{Act}\cup\{\inta\}} \sum_{e_a{:} E_a} c_a'(d',e_a) \rightarrow a(f_a'(d', e_a)) \cdot Y(g_a'(d', e_a)) \text{.}$$

\noindent The LPEs $X$ and $Y$ are divergence-preserving branching bismilar if there is a \textit{state mapping} $h: D \to D'$, a \textit{focus condition} $\textit{FC} : D \to \mathbb{B}$, a well founded ordering $<_M$ on $D$, and a cone labeling $p : D' \to \{\Delta, \nabla\}$ such that $h(d_0) = d_0'$ and the following requirements hold for all states $d{:}D$ in which invariant $\mathcal{I}$ holds:

\begin{enumerate}[label = \normalfont{\Roman{*}}]
    \item \label{req:bbsim_cones_converge_to_fc} If not a focus point, there is at least one internal step such that the target state is closer to the focus point:
    $$(\neg \textit{FC}(d)) \Rightarrow (\exists e_{\inta}{:}E_{\inta}.~ c_{\inta}(d, e_{\inta}) \wedge g_{\inta}(d, e_{\inta}) <_M d)$$
    
    \item \label{req:bbsim_closed_under_internal_actions} For every internal step, the mapping $h$ maps source and target states to the same states in the specification:
    $$\forall e_{\inta}{:} E_{\inta}.c_{\inta}(d,e_{\inta}) \Rightarrow h(d) = h(g_{\inta}(d, e_{\inta})) $$
    
    \item \label{req:bbsim_focus_points_mimic_specification_states} Every visible action in the specification must be enabled after a finite number of \inta actions for each corresponding focus point: For all $a {:} \textit{Act}$
    $$\forall e_a{:}E_a. (\textit{FC}(d) \wedge c_a'(h(d), e_a)) \Rightarrow (\exists d_{\inta}{:}D. d {\xrightarrow{\scriptsize{\inta}}}\!^* d_{\inta} \wedge c_a(d_{\inta},e_a))$$
    
    \item \label{req:bbsim_act_in_cone_to_spec} Every visible action in the implementation must be mimicked in the corresponding state in the specification: For all $a {:} \textit{Act}$
    $$\forall e_a{:}E_a.c_a(d, e_a) \Rightarrow c_a'(h(d), e_a)$$
    
    \item \label{req:bbsim_arguments_match} Matching actions have matching parameters: For all $a {:} \textit{Act}$
    $$\forall e_a{:}E_a. c_a(d, e_a) \Rightarrow f_a(d, e_a) = f'_a(h(d),e_a)$$
    
    \item \label{req::bbsim_endpoints_are_related} For all matching actions in specification and implementation, their endpoints must be related: For all $a {:} \textit{Act}$
    $$\forall e_a {:} E_a.c_a(d, e_a) \Rightarrow h (g_a(d,e_a)) = g_a'(h(d), e_a)$$
\end{enumerate}
\begin{enumerate}[label = \normalfont{\Roman{*}}$\!_\Delta$]
    \item \label{req:dpbb_int_loop} Any internal action in the specification is part of an $\inta$-loop:
    \[\forall_{e_{\inta}}{:}E_{\inta}.c_{\inta}'(h(d), e_{\inta}) \Rightarrow g_{\inta}'(h(d), e_{\inta}) = h(d)\]
    \item \label{req:dpbb_cones_loop} The cone labeling indicates whether or not a specification state allows an \inta-loop:
    $$p(h(d)) = \Delta \Leftrightarrow (\exists e_{\inta}{:} E_{\inta}.~ c'_{\inta}(h(d),e_{\inta}))$$
    \item \label{req:dpbb_focus_points_diverge} A cone is labelled as divergent if and only if it is possible to take an internal action in its focus points:
    $$\textit{FC}(d) \Rightarrow( p( h(d)) = \Delta \Leftrightarrow \exists e_{\inta}{:}E_{\inta}. c_{\inta}(d,e_{\inta}))$$
    \item \label{req:dpbb_non_divergent_cones_strictly_converge} All internal transitions within a non-divergent cone must bring us closer to a focus point:
    $$\forall e_{\inta}{:}E_{\inta}. ((p(h(d)) = \nabla \wedge c_{\inta}(d,e_{\inta})) \Rightarrow g_{\inta}(d,e_{\inta}) <_M d)$$
\end{enumerate}
\end{theorem}

\noindent \textit{Proof.} We define $R \subseteq D \times D'$ as  $\{\langle d,h(d) \rangle~|~ d \in D \wedge \mathcal{I}(d)\}$.

Proving that $R$ is a branching bisimulation, i.e.\ proving conditions $\textit{B}_1$ and $\textit{B}_2$ from Defininition \ref{def:dpbb}, follows the same general proof structure as is used in both \cite{Fokkink}, and \cite{Grootespringintveld}. We give a concise proof sketch.

\textbf{Condition} $\textit{B}_1$. Consider the states $d,d'{:}D$, and label $l{:}\textit{Act}\cup\{\inta\}$ such that $d\overset{a}{\to}d'$. As per Requirement \ref{req:bbsim_closed_under_internal_actions}, if $l=\inta$ then $h(d) = h(d')$. If $l\neq \inta$, then we have $h(d)\overset{l}{\to}h(d')$ as per Requirement \ref{req:bbsim_act_in_cone_to_spec}, and \ref{req::bbsim_endpoints_are_related}.

\textbf{Condition} $\textit{B}_2$. Consider the states $d{:}D, d'_2{:}D'$, and label $l{:}\textit{Act}\cup\{\inta\}$ such that $h(d) \overset{l}{\to} d_2'$.
If $l=\inta$, then $h(d_2') = h(d)$, as per Requirement \ref{req:dpbb_int_loop}.
If $l\neq\inta$, then there is a state $d_2{:}D$ such that $d \xrightarrow{\small{\inta}}\!\!\!^* d_2$ and $\textit{FC}(d_2)$ as per Requirement \ref{req:bbsim_cones_converge_to_fc} and $<_M$ being well founded. As per Requirements \ref{req:bbsim_focus_points_mimic_specification_states} and \ref{req::bbsim_endpoints_are_related}, there are states $d_3, d_4{:}D$ such that $d \xrightarrow{\small{\inta}}\!\!\!^* d_2 \xrightarrow{\small{\inta}}\!\!\!^* d_3 \overset{l}{\to} d_4$ and $h(d_4) = d_2'$.
From Requirement \ref{req:bbsim_closed_under_internal_actions} follows that all states along the $\inta$ path are related to $h(d)$.

We show that the branching bisimulation $R$ is also divergence-preserving by proving the two remaining conditions.

\textbf{Condition} $\textit{D}_1$. Consider the pair $\langle d, h(d)\rangle \in R$ and an infinite sequence $(d_n)_{n\in\Nat}$ over states in $D$ such that $d_0 = d$ and for any $n\in\Nat$ we have $h(d_n) = h(d)$ and $d_n{\xrightarrow{\scriptsize{\inta}}} d_{n+1}$.
We show that there is some $e_{\inta}{:}E_{\inta}$ such that $c'_{\inta}(h(d),e_{\inta})$ and $g'_{\inta}(h(d), e_{\inta}) = h(d)$.
If $h(d)$ is labeled $\Delta$ then this directly follows from Requirements \ref{req:dpbb_int_loop}, and \ref{req:dpbb_cones_loop}.

Assume, for sake of contradiction, that $h(d)$ is labeled as $\nabla$ instead. 
Since $<_M$ is a well-founded ordering on $D$, the sequence $(d_n)_{n\in\Nat}$ contains some minimal element $d_\bot$ such that no other element in the sequence is smaller than $d_\bot$.
However, as per Requirement \ref{req:dpbb_non_divergent_cones_strictly_converge}, any outgoing \inta action from $d_\bot$ must have an endpoint that is smaller than $d_\bot$, and thus the state that comes directly after $d_\bot$ in the sequence would have to be smaller, contradicting that $d_\bot$ is the minimal element.

\textbf{Condition} $\textit{D}_2$. Consider the pair $\langle d, h(d)\rangle \in R$ and an infinite sequence $(d'_n)_{n\in\Nat}$ over states in $D'$ such that $d'_0 = h(d)$ and for any $n\in\Nat$ we have $d~R~d'_n$, i.e.\ $d'_n = h(d)$, and $d'_n {\xrightarrow{\scriptsize{\inta}}} d'_{n+1}$.

Since $h(d)$ allows an $\texttt{int}$-loop, we have $p(h(d)) = \Delta$ as per Requirement \ref{req:dpbb_cones_loop}.
If $d$ is not a focus point then this action is enabled as per Requirement \ref{req:bbsim_cones_converge_to_fc}.
If $d$ is a focus point then this action is enabled as per Requirement \ref{req:dpbb_focus_points_diverge}, since its corresponding cone is labeled as $\Delta$. 
Requirement \ref{req:bbsim_closed_under_internal_actions} gives us that the endpoint of this internal action is related to $h(d)$.
Thus, if a state in the specification diverges then so do the related states in the implementation.

We thus conclude that the relation $R$ is a divergence-preserving branching bisimulation.\qed

\section{Models of the specification and the implementation} \label{sec:models_of_busyforbidden}
We now discuss the models of the specification and implementation of the busy-forbidden protocol, such that we can use the extended proof framework to prove them equivalent in Section \ref{sec:equivalence_proof}.
From here on, we use $N$ to denote the number of concurrent threads and we define $P = \{ p_1, \hdots, p_N \}$ to be the set containing all $N$ threads.

The linear process equation of the external behavior of the busy-forbidden protocol is given in Table \ref{spec:busy_forbidden_spec} in the appendix \cite{arxivVer}. 
The set $S$ contains the nodes shown in Figure \ref{fig:external-busy-forbidden}.
Each state in the specification is represented using a mapping $s$ that maps each thread to its current node, with each thread starting in the \textit{Free} node.
The set of specification states for $N$ threads is denoted by $D'_N$.
Note that each condition in the specification is the same as the conditions shown in Figure \ref{fig:external-busy-forbidden}.
The \textit{improbable} actions are considered to be \inta actions.

The linear process equation of the implementation is given in Table \ref{spec:busy_forbidden_impl} in the appendix \cite{arxivVer}.
All non-\texttt{typewriter} font actions are considered to be $\tau$ actions and \textit{italicized} actions are specifically considered to be \inta actions.
The set of implementation states $D_N$ is given in Definition \ref{def:states_impl}.
A part of each state consists of $N$ substates, with each substate giving the state of that specific thread.
The set of substates is given in Definition \ref{def:substates}, in which substates corresponding to the same node are grouped together.

\newcommand{\stated}{d = \langle d_{p_1}, d_{p_2}, \hdots, d_{p_N}, \textit{busy}, \textit{forbidden}, \textit{mtx}\rangle}
\newcommand{\statedtwo}{d' = \langle d_{p_1}', d_{p_2}', \hdots, d_{p_N}', \textit{busy}', \textit{forbidden}', \textit{mtx}' \rangle}

\begin{definition} \label{def:states_impl}
\normalfont{Each state in the linearized process of the busy-forbidden implementation for $N$ threads is defined as the tuple 
\[\stated {:} D_N\text{, in which:}\]
\begin{itemize}
    \item $d_{p_1}, d_{p_2}, \hdots, d_{p_N}$ are the substates of threads $1$ through $N$.
    \item $\textit{busy} : P \rightarrow \mathbb{B}$ is the mapping that keeps track of all the busy flags, in which $\textit{busy}(p)$ is the current value of the busy flag of thread $p$.
    \item $\textit{forbidden} : P \rightarrow \mathbb{B}$ is the mapping that keeps track of all the forbidden flags in the same way as the $\textit{busy}$ mapping.
    \item $\textit{mtx}$ is a boolean that indicates whether the mutual exclusion variable $\textit{mtx}$ is locked or unlocked.
\end{itemize}}
\end{definition}

\begin{definition} \label{def:substates}
    \normalfont{The set of substates for each individual process is defined as the union of the following sets:
    \begin{itemize}
        \item $\textit{Free} = \{\textit{Free}\}$, $\textit{ES} = \{\textit{ES}_1, \textit{ES}_2, \textit{ES}_3, \textit{ES}_4 \}$, $\textit{LOE} = \{\textit{LOE}\}$,\\
        $\textit{Shared} = \{ \textit{Shared} \}$, $\textit{LS} = \{ \textit{LS}_1, \textit{LS}_2\}$, $\textit{EE} = \{ \textit{EE} \}$, $\textit{LOS} = \{ \textit{LOS}_1, \textit{LOS}_2 \}$,
        $\textit{Exclusive} = \{ \textit{Exclusive} \}$, and $\textit{OE} = \{ \textit{OE}_1, \textit{OE}_2 \}$,
        \item $\textit{SAF} = \{\textit{SAF}_U | U \subset P \}\! \cup\!\{ \textit{SAF}_{p_x,U} | p_x{:}P, U\subset P\}\!\cup\!\{ \textit{SAF}^{\textit{undo}}_{p_x, U} | p_x{:}P, U\subset P \}$,
        \item and $\textit{LE} = \{ \textit{LE}_U | U \subseteq P \wedge U \neq \emptyset\}$.
    \end{itemize}
    Note that the singleton sets, such as \textit{Free}, contain a single state with the same name as the set and do not contain themselves.}
\end{definition}

In the initial state of the implementation for $N$ threads, all substates are set to \textit{Free}, \textit{busy} and \textit{forbidden} map each thread $p$ to $\false$ and \textit{mtx} is set to $\false$.

Since the state tuple contains a large number of elements, we use a shorthand notation for writing down the resulting state.
All elements which remain the same are not listed and are abbreviated with ``$\textit{etc}.$".
A substate or the \textit{mtx} variable being changed in the resulting state is denoted with the ``$=$" operator, where the lefthand side is assigned the value on the righthand side, e.g.\ $d_p = \textit{ES}_2$ indicates that the substate of thread $p$ becomes $\textit{ES}_2$ in the next state.
The function update $f[ e \mapsto n]$ specifies that in the next state $f(x)$ equals the new value $n$ if $x\approx e$ and otherwise equals its original value.

We introduce the Invariants \ref{inv:exclusive_exclusive}, \ref{inv:busy_in_shared}, and \ref{inv:forbidden}. 
These exclude some unreachable states and show that for any given state, the exact values of \textit{busy}, \textit{forbidden}, and \textit{mtx} can be inferred from just the set of substates, i.e.\ $d_{p_1}, d_{p_2}, \hdots, d_{p_N}$.
In the proof of Invariant \ref{inv:exclusive_exclusive}, we show that the value of \textit{mtx} can be inferred from just the set of substates and that it is not possible to have multiple threads simultaneously present in the set of states fenced off by the mutex operations.
We show that the values of the \textit{busy} and \textit{forbidden} flags can also be inferred from just the set of substates in the proofs of Invariants \ref{inv:busy_in_shared} and \ref{inv:forbidden}.

The exact proofs for these invariants can be found in the appendix \cite{arxivVer}.
All of them follow the same general structure.
Namely, the actions that result in a thread entering or leaving the given set of states, e.g.\ $B$, are the exact same actions that result in the value, e.g.\ $\textit{busy}(p)$, being altered.
And thus the exact values can be inferred from just the set of substates.

\begin{invariant} \label{inv:exclusive_exclusive}
The following invariant holds in the initial state and all subsequent states of the implementation: Given any state $d{:}D$ as per Definition \ref{def:states_impl},
$$\exists p {:} P.~ d_p \in M \Leftrightarrow \textit{mtx} \text{, and }\forall p_x, p_y {:} P .~ d_{p_x},d_{p_y} \in M \Rightarrow p_x = p_y \text{,}$$
where $M =\textit{SAF} \cup \textit{LOS} \cup \textit{Exclusive} \cup \textit{LE} \cup \{ \textit{OE}_2\}$.
\end{invariant}

\begin{invariant} \label{inv:busy_in_shared}
The following invariant holds in the initial state and all subsequent states of the implementation: Given any state $d{:}D$ as per Definition \ref{def:states_impl},
$$\forall p{:}P. d_p \in B \Leftrightarrow \textit{busy}(p) \text{, where } B = \textit{LOE} \cup \textit{Shared} \cup \{ \textit{ES}_1,\textit{ES}_4, \textit{LS}_2\} \text{.}$$
\end{invariant}

\begin{invariant} \label{inv:forbidden}
The following invariant holds in the initial state and all subsequent states of the implementation: Given any state $d{:}D$ as per Definition \ref{def:states_impl},
$$\forall p{:}P. \textit{forbidden}(p) \Longleftrightarrow \exists q{:}P. d_q \in F\!\text{,}$$
where $F = \textit{LOS} \cup \textit{Exclusive} \cup \{\textit{LE}_U | U \subset P \wedge p\in U\} \cup \{\textit{SAF}_U | U \subset P \wedge p \in U\} \cup \{\textit{SAF}_{p,U} | U \subset P \} \cup \{\textit{SAF}_{p,U}^{\textit{undo}} | U \subset P\}$.
\end{invariant}

\section{Correctness of the busy-forbidden protocol} \label{sec:equivalence_proof}
The state mapping, focus condition, state ordering and cone labeling used during the equivalence proof are given in Definitions \ref{def:statemapping}, \ref{def:focuscondition}, \ref{def:order}, and \ref{def:cone_labeling}, respectively.
These data objects only need to use substates since the values of the \textit{busy}, \textit{forbidden}, and \textit{mtx} data objects can be directly inferred from the substates in any given state.
\begin{definition} \label{def:statemapping}
\normalfont{We define our state-mapping $h : D_N \to D_N'$ as follows:
    $$h(\langle d_1, d_2, \hdots, d_N, \textit{busy}, \textit{forbidden}, \textit{mtx}\rangle) = s \text{ where } s(p) = h_P(d_p) \text{ for any } p {:} P\text{.}$$
The mapping $h_P$, referred to as the substate-mapping, maps each substate to the specification state with the same name as the set, shown in Definition \ref{def:substates}, that it belongs to, e.g. $h_P(\textit{ES}_3) = \textit{ES}$ and $h_P(\textit{SAF}_{\{p_1, p_3, p_4\}}) = \textit{SAF}$.}
\end{definition}

\begin{definition} \label{def:focuscondition}
\normalfont{We define our focus condition $\textit{FC} : D_N \to \mathbb{B}$ as follows:
    $$\textit{FC}(\langle d_{p_1}, d_{p_2}, \hdots, d_{p_N}, \textit{busy, forbidden, mtx}\rangle) = \bigwedge_{p_x{:}P} \textit{FC}_{p_x}(d_{p_x}) \text{,}$$
where $\textit{FC}_{p_x}(d_{p_x}) \overset{\scriptsize{\textit{def}}}{=} p_x \in \{ \textit{Free}, \textit{ES}_1, \textit{LOE, Shared}, \textit{LS}_1, \textit{EE}, \textit{SAF}_\emptyset, \textit{LOS}_1, \textit{Exclu-}\\\textit{sive}, \textit{LE}_{\{p_x\}}, \textit{OE}_1\}$.
We refer to the predicate $\textit{FC}_{p_x}$, for any given $p_x{:}P$, as the sub-focus condition.}
\end{definition}

\begin{definition} \label{def:order}
\normalfont{Given two states $\stated$ and $d' = \langle d'_{p1}, d'_{p2}, \hdots,$ $ d'_{pN}, \textit{busy}', \textit{forbidden}', \textit{mtx}' \rangle$, we define the ordering on these states as follows:
    $$d <_M d' \overset{\textit{def}}{=} \bigwedge_{p{:}P} d_p <_p d_p' \text{,}$$
where, given some thread $p{:}P$, the ordering $<_p$ on its substates is defined such that only the following holds:
\begin{itemize}
    \item $\textit{ES}_1 <_p \textit{ES}_2 <_p \textit{ES}_3 <_p \textit{ES}_4 $, 
    $\textit{LS}_1 <_p \textit{LS}_2$, 
    $\textit{LOS}_1 <_p \textit{LOS}_2$,\\
    and $\textit{OE}_1 <_p \textit{OE}_2$,
    \item $\textit{SAF}_{p_x,U} <_p \textit{SAF}_U$ iff $p_x \in U$ for any given $U{:}\mathcal{P}(P)$ and $p_x{:}P$,\\
    $\textit{SAF}_{U\setminus\{p_x\}} <_p \textit{SAF}_{p_x,U}$ for any given $U{:}\mathcal{P}(P)$ and $p_x{:}P$,\\
    $\textit{SAF}_U < \textit{SAF}^{\textit{undo}}_{p_x,U'}$ for any given given $U,U'{:}\mathcal{P}(P)$ and $p_x{:}P$,
    \item $\textit{LE}_U <_p \textit{LE}_{U'}$ iff $U \subset U' \wedge p \in U$ or $p \in U \wedge p\not\in U'$ for any given $U,U'{:}\mathcal{P}(P)$
\end{itemize}}
\end{definition}

\begin{definition} \label{def:cone_labeling}
    \normalfont{We define the cone labeling $p : D'_N \to \{\Delta, \nabla \}$ as follows:
    Given any state $s {:} D_N'$, $p(s) = \Delta$ iff $\exists q{:}P. s(q) \in \{ \textit{SAF}, \textit{LE}\} \vee (\exists q:P. s(q) = \textit{ES} \wedge \exists q'{:}P. q' \in \{\textit{LOS}, \textit{Exclusive}\})$
    otherwise $p(s) = \nabla$.}
\end{definition}

The specification indicates that if there is one thread in the \textit{ES} node and one thread in the \textit{SAF} node, either one of them should be able to progress to the next node.
This is not simultaneously possible in the implementation, as progressing to the \textit{LOE} node requires the \textit{busy} flag to be $\true$ and the \textit{forbidden} flag to be $\false$, while progressing to the \textit{LOS} node requires all \textit{busy} flags to be false and all \textit{forbidden} flags to be $\true$.
Thus, the subfocus point of each node is chosen such that the external actions are enabled directly given that they would also be enabled in the specification, with the exception of $\textit{SAF}_\emptyset$ which is used as the focus point of the \textit{SAF} node.

We show that there is a path of \inta actions from this to some state $d_{\inta}$ in which the transition to \textit{LOE} is enabled.
This is outlined in Theorem \ref{thm:aux_equiv} for which the proof is given in the appendix \cite{arxivVer}.
The general idea behind the proof is that if the \textit{forbidden} flag is set before it is read by the thread in the \textit{ES} node, the \textit{busy} flag will be set back to false.
Repeating this, leads to all \textit{busy} flags being $\false$ and all \textit{forbidden} flags being $\true$, thus enabling the transition to \textit{LOE}.

We now conclude by proving the implementation and specification of the busy-forbidden protocol equivalent in Theorem \ref{thm:busyforbidden_equiv}.

\begin{theorem} \label{thm:aux_equiv}
    \normalfont{Given some state $d{:}D$, some thread $p_{\textit{SAF}}{:}P$, and some data configuration $e_\tau{:}E_\tau$ such that $\textit{FC}(d)$ and $c'_\tau(h(d), e_\tau)$ hold, $h(d)(p_{\textit{SAF}}) = \textit{SAF}$ and $g'_\tau(h(d), e_\tau) = \textit{LOE}$.
    There must be some state $d_{\inta}{:}D$ such that $d {\xrightarrow{\scriptsize{\inta}}}\!^* d_{\inta}$ and $c_\tau(d_{\inta}, e_\tau)$ hold and $h(g_\tau(d_{\inta}, e_\tau))(p_{\textit{SAF}}) = \textit{LOE}$.}
\end{theorem}

\begin{theorem} \label{thm:busyforbidden_equiv}
    \normalfont{The LPE of the implementation given in Table \ref{spec:busy_forbidden_impl} and the LPE of the specification given in Table \ref{spec:busy_forbidden_spec} are divergence-preserving branching bisimilar.}
\end{theorem}
\noindent \textit{Proof.} To prove the aforementioned equivalence, we show that all ten requirements given in Theorem \ref{thm:new_cnf} hold using Invariants \ref{inv:exclusive_exclusive}, \ref{inv:busy_in_shared}, and \ref{inv:forbidden}, and the state mapping, focus condition, ordering and cone labeling, given in Definitions \ref{def:statemapping}, \ref{def:focuscondition}, \ref{def:order}, and \ref{def:cone_labeling}, respectively.
From the linear process equation, it is relatively easy to see that Requirements \ref{req:bbsim_cones_converge_to_fc}, \ref{req:bbsim_closed_under_internal_actions}, \ref{req:bbsim_arguments_match}, \ref{req::bbsim_endpoints_are_related}, \ref{req:dpbb_int_loop}, and \ref{req:dpbb_cones_loop} are not invalidated.
As such, we refer the reader to their extended proofs, found in the appendix \cite{arxivVer}.

Both the implementation and specification contain exactly three externally observable actions that are not always enabled.
For these actions, we show that if the action in the specification is enabled, the same action is also enabled in the corresponding focus point in the implementation, and if the action in the implementation is enabled, the corresponding specification action is also enabled,
thus showing that Requirements \ref{req:bbsim_focus_points_mimic_specification_states}, and \ref{req:bbsim_act_in_cone_to_spec} hold.

The first action is the $\text{load}(\textit{Forbidden}(p),\false, p)$ action in $\textit{ES}_2$ and the $\tau$ transition from the \textit{ES} to the \textit{LOE} node in the specification.
The \textit{load} action is only enabled when $\textit{forbidden}(p)$ is $\false$, and the $\tau$ transition in the specification is only enabled if there are no threads in \textit{LOS} or \textit{Exclusive} node.
As per Invariant \ref{inv:forbidden}, these conditions hold exactly when they hold in the corresponding focus points.

The second action is the $\text{lock}(p)$ action in \textit{EE} and the $\tau$ transition in the \textit{EE} node in the specification.
The \textit{lock} action is only enabled when \textit{mtx} is $\false$, and the $\tau$ transition in the specification is only enabled if there is no thread in the \textit{SAF}, \textit{LOS}, \textit{Exclusive}, and \textit{LE} node.
As per Invariant \ref{inv:exclusive_exclusive}, these conditions, again, hold exactly when they would hold in the corresponding focus points of the implementation.

The third action is the $\text{load}(\textit{Busy}(p_x),\false, p)$ action in $\textit{SAF}_{p_x,U}$ and the $\tau$ transition from the \textit{SAF} to the \textit{LOS} node in the specification.
The \textit{load} action is only enabled when $\textit{Busy}(p)$ is $\false$ and the $\tau$ transition is only enabled if there is no thread in the \textit{LOE} and \textit{Shared} nodes.
As per Invariant \ref{inv:busy_in_shared}, if $\textit{busy}(p)$ is $\false$ then the \textit{LOE} and \textit{Shared} node are empty and thus, if the action is enabled in the implementation, it is also enabled in the specification.
As per the same invariant, the only focus points in which the action would not be enabled while it would be in the corresponding specifications state, are the ones in which a thread is in the \textit{SAF} node, i.e.\ some thread $p{:}P$ has the substate $\textit{SAF}_\emptyset$.
In these cases, as per Theorem \ref{thm:aux_equiv}, there must be some finite path of \inta actions to some state $d_{\inta}$ in which this action is enabled.

In the corresponding focus points for the \textit{SAF} and \textit{LE} cone, there is always at least one internal action enabled.
In the focus point for the \textit{ES} cone, the $\textit{load}(\textit{Forbidden}(p), \true, p)$ action is enabled iff $\textit{forbidden}(p)$ is $\true$.
As per Invariant \ref{inv:forbidden}, the only focus points in which $\textit{Forbidden}(p)$ is $\true$ are the ones in which the \textit{LOS} or \textit{Exclusive} node are occupied.
In all other focus points, there are no further internal actions enabled.
Thus Requirement \ref{req:dpbb_focus_points_diverge} holds.

If a cone is labelled as non-diverging ($\nabla$), then each thread should be in one of the following nodes:
\textit{Free}, \textit{LOE}, \textit{Shared}, \textit{LS}, \textit{EE}, \textit{LOS}, \textit{Exclusive}, or \textit{OE}, or \textit{ES}, given that there are no threads present in either \textit{LOS} or \textit{Exclusive}.
With the exception of the $\textit{load}(\textit{Forbidden}(p), \true, p)$ action in the \textit{ES} node, all the internal actions within these nodes take us closer to a focus point.
As per Invariant \ref{inv:forbidden}, \textit{forbidden} is $\true$ only if there is a thread present in either the \textit{LOS} or \textit{Exclusive}, \textit{LE}, or \textit{SAF} node, which are known to be empty.
Thus Requirement \ref{req:dpbb_non_divergent_cones_strictly_converge} also holds and the implementation and specification are divergence-preserving branching bisimilar as per Theorem \ref{thm:new_cnf}. \qed

\section{Conclusion and future work}
We have extended the cones and foci proof framework \cite{Fokkink, Grootespringintveld} with four additional requirements, i.e.\ Requirements \ref{req:dpbb_int_loop}, \ref{req:dpbb_cones_loop}, \ref{req:dpbb_focus_points_diverge}, and \ref{req:dpbb_non_divergent_cones_strictly_converge}, such that it can be used to prove divergence-preserving branching bisimulation.
We have proven this extension to be sound and have used it to prove the implementation and specification of the novel busy-forbidden protocol \cite{paraterm} to be equivalent.

We note some opportunities to extend upon the work in this paper:
\begin{itemize}
    \item The completeness of the extended cones and foci proof framework has not been formally proven. We assume its completeness due to the weakening of Requirement \ref{req:bbsim_focus_points_mimic_specification_states}, and it is of similar interest as to whether this Requirement can be made stronger without loss of our assumed completeness.
    \item As mentioned before, the original cones and foci proof framework has been used for the verification of the sliding window protocol \cite{slidingwindow}. 
    The communication channels used by this protocol are unreliable and thus allow divergence.
    As such, the sliding window protocol could provide an interesting case study for our extension of the cones and foci proof framework.
    \item The diverging loops in the external behavior are considered to be \textit{improbable}, as such, we abstract away any actual, but potentially informative, probabilistic analysis of the protocol.
\end{itemize}
\printbibliography 

\newpage
\appendix
\section{Process equations of the original implementation model} \label{appendix:original_model}
Tables \ref{spec:enter_shared}, \ref{spec:leave_shared}, \ref{spec:enter_exclusive}, and \ref{spec:leave_exclusive} contain the process equations used to model a thread $p{:}P$ calling the \texttt{enter\_shared}, \texttt{leave\_shared}, \texttt{enter\_exclusive}, and \texttt{leave\_exclusive} functions, shown in Table \ref{table:busy_forbidden}, respectively.
Table \ref{spec:busy_forbidden_thread} contains the process equation of a thread $p{:}P$ which repeatedly attempts to enter and then leave either the shared or exclusive section by calling these functions.
Table \ref{spec:busy_forbidden_comm} contains the process equations for modelling the \textit{busy} and \textit{forbidden} flags and the mutex.
The initial composition of these processes is given in Table \ref{spec:busy_forbidden_impl}.

The $||$ operator models the parallel composition of processes and allows for both the interleaving of actions as-well as for actions to occur in parallel.
The \textbf{comm} and \textbf{allow} operators are used to enforce that only the combinations of actions, listed under the \textbf{comm} operation, can and must occur in parallel.
This enforces communication between the \textit{Thread} processes and the \textit{Flags} and \textit{Mutex} processes.

The linearization of the implementation, is shown in Table \ref{spec:busy_forbidden_impl}.
With the exception of this linearization, all process equations shown in Appendix \ref{appendix:original_model} are taken from the original paper \cite{paraterm}.

\begin{table}[]
    \centering
    \begin{tabular}{|p{\textwidth}|}\hline
    $\textbf{allow}(\{$\\
	\tab{}{1}$\textit{store}, \textit{load},$\\
	\tab{}{1}$\textit{lock}, \textit{unlock},$\\
	\tab{}{1}$\textit{internal, improbable,}$\\
	\tab{}{1}$\texttt{enter\_shared\_call}, \texttt{enter\_shared\_return},$\\
	\tab{}{1}$\texttt{leave\_shared\_call}, \texttt{leave\_shared\_return},$\\
	\tab{}{1}$\texttt{enter\_exclusive\_call}, \texttt{enter\_exclusive\_return},$\\
	\tab{}{1}$\texttt{leave\_exclusive\_call}, \texttt{leave\_exclusive\_return}$\\
	\tab{}{1}$\}, \textbf{comm}(\{$\\
	\tab{}{2}$\textit{store}_f | \textit{store}_p \rightarrow \textit{store},$\\
	\tab{}{2}$\textit{load}_f | \textit{load}_p \rightarrow \textit{load},$\\
	\tab{}{2}$\textit{lock}_m | \textit{lock}_p \rightarrow \textit{lock},$\\
	\tab{}{2}$\textit{unlock}_m | \textit{unlock}_p \rightarrow \textit{unlock}$\\
	\tab{}{2}$\}, $\\
    \tab{}{2}$\textit{Thread}(p_1)~||$\\
    \tab{}{3}$\vdots$\\
    \tab{}{2}$\textit{Thread}(p_N)~||$\\
    \tab{}{2}$\textit{Flags}(\lambda f{:}F.\false)~||$\\
    \tab{}{2}$\textit{Mutex}(\false)$\\
    \tab{}{2}$)$\\
    \tab{}{1}$)$\\ \hline
    \end{tabular}
\caption{Parallel composition used to model the busy-forbidden protocol.}
\label{spec:busy_forbidden_init}
\end{table}

\begin{table}[]
    \centering
    \begin{tabular}{|p{\textwidth}|} \hline
    $\textit{Flags}( \textit{flags} : F \rightarrow \mathbb{B}) =$\\
    $\sum_{f{:}F,p{:}P}. ($\\
    \tab{}{1}$\sum_{b{:}\mathbb{B}}. \textit{store}_f(f, b, p)~\cdot~\textit{Flag}(\textit{flags}[f\mapsto b])$\\
    \tab{+}{1}$\textit{load}_f(f, \textit{flags}(f), p)~\cdot~\textit{Flag}(\textit{flags})$\\
    \tab{}{1}$)$\\
    \\
    $\textit{Mutex}( \textit{locked} : \mathbb{B}) =$\\
    $\sum_{p{:}P}. ($\\
    \tab{}{1}$\textit{locked}$\\
    \tab{$\rightarrow$}{2}$\textit{lock}_m(p)~\cdot~\textit{Mutex}(\true)$\\
    \tab{$\diamond$}{2}$\textit{unlock}_m(p)~\cdot~\textit{Mutex}(\false)$\\
    \tab{}{1}$)$ \\ \hline
    \end{tabular}
    \caption{Model of the components used during busy-forbidden.}
\label{spec:busy_forbidden_comm}
\end{table}

\begin{table}[]
    \centering
    \begin{tabular}{|p{\textwidth}|} \hline
    $\textit{EnterShared}(p : P) =$\\
    \tab{}{1}$\texttt{enter\_shared\_call}(p)~\cdot$\\
    \tab{}{1}$\textit{TryBothFlags}(p) ~\cdot$\\
    \tab{}{1}$\texttt{enter\_shared\_return}(p)$\\
    \\
    $\textit{TryBothFlags}(p : P) =$\\
    \tab{}{1}$\textit{store}_p(\textit{Busy}(p), \false, p)~\cdot~($\\
    \tab{}{2}$\textit{load}_p(\textit{Forbidden}(p),~ \true,~ p)~\cdot$\\
    \tab{}{2}$\textit{store}_p(\textit{Busy}(p), \false, p)~\cdot~\textit{improbable}~\cdot~\textit{TryBothFlags}(p)$\\
    \tab{+}{2}$\textit{load}_p(\textit{Forbidden}(p), \false, p)$\\
    \tab{)}{2} \\ \hline
    \end{tabular}
    \caption{Model of the \texttt{enter\_shared} function shown in Table \ref{table:busy_forbidden}.}
\label{spec:enter_shared}
\end{table}

\begin{table}[]
    \centering
    \begin{tabular}{|p{\textwidth}|} \hline
    $\textit{LeaveShared}(p : P) =$\\
    \tab{}{1}$\texttt{leave\_shared\_call}(p)~\cdot$\\
    \tab{}{1}$\textit{store}_p(\textit{Busy}(p), \false, p)~\cdot$\\
    \tab{}{1}$\texttt{leave\_shared\_return}(p)$ \\ \hline
    \end{tabular}
    \caption{Model of the \texttt{leave\_shared} shown in Table \ref{table:busy_forbidden}.}
    \label{spec:leave_shared}
\end{table}

\begin{table}[]
    \centering
    \begin{tabular}{|p{\textwidth}|} \hline
    $\textit{EnterExclusive}(p : P) =$\\
    \tab{}{1}$\texttt{enter\_exclusive\_call}(p)$\\
    \tab{}{1}$\textit{lock}_p(p)~\cdot$\\
    \tab{}{1}$\textit{SetAllForbiddenFlags}(p,~ \emptyset)~\cdot$\\
    \tab{}{1}$\texttt{enter\_exclusive\_return}(p)$\\
    \\
    $\textit{SetAllForbiddenFlags}(p : P,~ \textit{forbidden} : Set(P)) =$\\
    \tab{}{1}$(\forall_{p'{:}P}. p \in \textit{forbidden})$\\
    \tab{$\rightarrow$}{2}$internal$\\
    \tab{$\diamond$}{2}$\sum_{p'{:}P}. \textit{store}_p(\textit{Forbidden}(p'), \true, p)~\cdot~($\\
    \tab{}{3}$\textit{load}_p(\textit{Busy}(p'), \false, p)~\cdot$\\
    \tab{}{3}$\textit{SetAllForbiddenFlags}(p,~ \textit{forbidden} \cup \{p'\})$\\
    \tab{+}{3}$l\textit{oad}_p(\textit{Busy}(p'), \true, p)~\cdot$\\
    \tab{}{3}$\textit{store}_p(\textit{Forbidden}(p'), \false, p)~\cdot~\textit{improbable}~\cdot$\\
    \tab{}{3}$\textit{SetAllForbiddenFlags}(p,~ \textit{forbidden} \setminus \{ p'\})$\\
    \tab{+}{3}$\textit{store}_p(\textit{Forbidden}(p'), \false, p)~\cdot~\textit{improbable}~\cdot$\\
    \tab{}{3}$\textit{SetAllForbiddenFlags}(p,~ \textit{forbidden} \setminus \{p'\})$\\
    \tab{}{3}$)$ \\ \hline
    \end{tabular}
    \caption{Model of the \texttt{enter\_exclusive} function shown in Table \ref{table:busy_forbidden}.}
    \label{spec:enter_exclusive}
\end{table}

\begin{table}[]
    \centering
    \begin{tabular}{|p{\textwidth}|} \hline
    $\textit{LeaveExclusive}(p : P) =$\\
    \tab{}{1}$\texttt{leave\_exclusive\_call}(p)~\cdot$\\
    \tab{}{1}$\textit{AllowAllThreads}(p,\emptyset)~\cdot$\\
    \tab{}{1}$\textit{unlock}_p(p)~\cdot$\\
    \tab{}{1}$\texttt{leave\_exclusive\_return}(p)$\\
    \\
    $\textit{AllowAllThreads}(p : P,~\textit{allowed} : \textit{Set}(P)) =$\\
    \tab{}{1}$(\forall_{q{:}P}. q \in \textit{allowed})$\\
    \tab{$\rightarrow$}{2}\textit{internal}\\
    \tab{$\diamond$}{2}$\sum_{p'{:}P}.($\\
    \tab{}{2}$\textit{store}_p(\textit{Forbidden}(p'), \false, p)~\cdot$\\
    \tab{}{2}$\textit{AllowAllThreads}(p,~\textit{allowed}\cup\{p'\})$\\
    \tab{+}{2}$\textit{store}_p(\textit{Forbidden}(p'), \true, p)~\cdot~\textit{improbable}~$\\
    \tab{}{2}$\textit{AllowAllThreads}(p,~\textit{allowed}\setminus\{p'\})$\\
    \tab{}{2}$)$ \\ \hline
    \end{tabular}
    \caption{Model of the \texttt{leave\_exclusive} function shown in Table \ref{table:busy_forbidden}.}
    \label{spec:leave_exclusive}
\end{table}

\begin{table}[]
    \centering
    \begin{tabular}{|p{\textwidth}|} \hline
    $\textit{Thread}(p : P)=$\\
    \tab{}{1}$\textit{EnterShared}(p)~\cdot$\\
    \tab{}{1}$\textit{LeaveShared}(p)~\cdot$\\
    \tab{}{1}$\textit{Thread}(p)$\\
    \tab{+}{1}$\textit{EnterExclusive}(p)~\cdot$\\
    \tab{}{1}$\textit{LeaveExclusive}(p)~\cdot$\\
    \tab{}{1}$\textit{Thread}(p)$ \\ \hline
    \end{tabular}
    \caption{Model of a thread \textit{p} interacting with the protocol.}
\label{spec:busy_forbidden_thread}
\end{table}

\begin{table}[]
    \centering
    \begin{tabular}{|p{\textwidth}|} \hline
    $\textit{BF}(s : P\rightarrow S) =$\\
    \tab{}{1}$\sum_{p{:}P} (s(p) \approx \textit{Free}) \to \texttt{enter\_share\_call}(p) \cdot \textit{BF}(s[p\mapsto \textit{ES}])$\\
    \tab{$+$}{1}$\sum_{p{:}P} (s(p) \approx \textit{ES} \wedge \neg\exists p'{:}P.s(p) \in \{\textit{LOS, Exclusive}\}) \to \textit{improbable} \cdot \textit{BF}(s)$\\
    \tab{$+$}{1}$\sum_{p{:}P} (s(p) \approx \textit{ES} \wedge \exists p'{:}P.s(p) \in \{\textit{LOS, Exclusive}\}) \to \tau \cdot \textit{BF}(s[p\mapsto\textit{LOE}])$\\
    \tab{$+$}{1}$\sum_{p{:}P} (s(p) \approx \textit{LOE}) \to \texttt{enter\_shared\_return}(p) \cdot \textit{BF}(s[p\mapsto\textit{Shared}])$\\
    \tab{$+$}{1}$\sum_{p{:}P} (s(p) \approx \textit{Shared}) \to \texttt{leave\_shared\_call}(p) \cdot \textit{BF}(s[p\mapsto\textit{LS}])$\\
    \tab{$+$}{1}$\sum_{p{:}P} (s(p) \approx \textit{LS}) \to \texttt{leave\_shared\_return}(p) \cdot \textit{BF}(s[p\mapsto\textit{Free}])$\\
    \\
    \tab{$+$}{1}$\sum_{p{:}P} (s(p) \approx \textit{Free}) \to \texttt{enter\_exclusive\_call}(p) \cdot \textit{BF}(s[p\mapsto\textit{EE}])$\\
    \tab{$+$}{1}$\sum_{p{:}P} (s(p) \approx \textit{EE} \wedge \neg\exists p'{:}P. s(p') \in \{\textit{SAF, LOS, Exclusive}\}) \to \tau \cdot \textit{BF}(s[p\mapsto\textit{SAF}])$\\
    \tab{$+$}{1}$\sum_{p{:}P} (s(p) \approx \textit{SAF}) \to \textit{improbable}\cdot\textit{BF}(s)$\\
    \tab{$+$}{1}$\sum_{p{:}P} (s(p) \approx \textit{SAF} \wedge \neg\exists p'{:}P. s(p) \in \{\textit{LOE, Shared}\}) \to \tau \cdot \textit{BF}(s[p\mapsto \textit{LOS}])$\\
    \tab{$+$}{1}$\sum_{p{:}P} (s(p) \approx \textit{LOS}) \to \texttt{enter\_exclusive\_return}(p) \cdot \textit{BF}(s[p\mapsto\textit{Exclusive}])$\\
    \tab{$+$}{1}$\sum_{p{:}P} (s(p) \approx \textit{Exclusive}) \to \texttt{leave\_exclusive\_call}(p) \cdot \textit{BF}(s[p\mapsto\textit{LE}])$\\
    \tab{$+$}{1}$\sum_{p{:}P} (s(p) \approx \textit{LE}) \to \textit{improbable} \cdot \textit{BF}(s)$\\
    \tab{$+$}{1}$\sum_{p{:}P} (s(p) \approx \textit{LE}) \to \tau \cdot \textit{BF}(s[p\mapsto\textit{OE}])$\\
    \tab{$+$}{1}$\sum_{p{:}P} (s(p) \approx \textit{OE}) \to \texttt{leave\_exclusive\_return}(p)\cdot\textit{BF}(s[p\mapsto\textit{Free}])$ \\ \hline
    \end{tabular}
    \caption{Process equation of the external behavior as shown in Figure \ref{fig:external-busy-forbidden}.}
\label{spec:busy_forbidden_spec}
\end{table}

\begin{table}
    \centering
    \begin{tabular}{|p{\textwidth}|} \hline
    $\textit{BF}(d=\langle d_{p_1}, \hdots, d_{p_N}, \textit{busy}, \textit{forbidden}, \textit{mutex}\rangle : D_N) =$\\
    \tab{}{1}$\sum_{p{:}P} (d_p \approx \textit{Free}) \to \texttt{enter\_shared\_call}(p) \cdot \textit{BF}(\langle d_p = \textit{ES}_2 , \etc)$\\
    \tab{+}{1}$\sump (d_p \approx \textit{Free}) \to \texttt{enter\_exclusive\_call}(p) \cdot \textit{BF}(\langle d_p = \textit{EE}, \etc)$\\ 
\hdashline
    \tab{+}{1}$\sump (d_p \approx \textit{ES}_2) \to \textit{store}(\textit{Busy}(p), \true, p) \cdot \textit{BF}(\langle d_p = \textit{ES}_1,\etc)$\\
    \tab{+}{1}$\sump (d_p \approx \textit{ES}_1 \wedge \textit{forbidden}(p))$\\
    \tab{$\to$}{2}$\textit{load}(\textit{Forbidden}(p), \true, p) \cdot \textit{BF}(\langle d_p = \textit{ES}_4, \etc)$\\
    \tab{+}{1}$\sump (d_p \approx \textit{ES}_1 \wedge \neg\textit{forbidden}(p))$\\
    \tab{$\to$}{2}$\text{load}(\textit{Forbidden}(p), \false, p) \cdot \textit{BF}(\langle d_p = \textit{LOE}, \etc)$\\
    \tab{$+$}{1}$\sump (d_p \approx \textit{ES}_4) \to \textit{store}(\textit{Busy}(p), \false, p) \cdot \textit{BF}(\langle d_p = \textit{ES}_3, \textit{busy}[p \mapsto \false], \etc \rangle$\\
    \tab{$+$}{1}$\sump (d_p \approx \textit{ES}_3) \to \textit{improbable} \cdot \textit{BF}(\langle d_p = \textit{ES}_2, \etc)$\\
\hdashline
    \tab{+}{1}$\sump (d_p \approx \textit{LOE}) \to \texttt{enter\_shared\_return}(p) \cdot \textit{BF}(\langle d_p = \textit{Shared}, \etc)$\\
\hdashline
    \tab{+}{1}$\sump (d_p \approx \textit{Shared}) \to \texttt{leave\_shared\_call}(p) \cdot \textit{BF}(\langle d_p = \textit{LS}_2, \etc)$\\
\hdashline
    \tab{+}{1}$\sump (d_p \approx \textit{LS}_2) \to \textit{store}(\textit{Busy}(p), \false, p) \cdot \textit{BF}(\langle d_p = \textit{LS}_1, \textit{busy}[p \mapsto \false],\etc)$\\
    \tab{+}{1}$\sump (d_p \approx \textit{LS}_1) \to \texttt{leave\_shared\_return}(p) \cdot \textit{BF}(\langle d_p = \textit{EE}, \etc)$\\
\hdashline
    \tab{+}{1}$\sump (d_p \approx \textit{EE} \wedge \neg\textit{mutex}) \to \text{lock}(p) \cdot \textit{BF}(\langle d_p = \textit{SAF}_\emptyset, \textit{mtx} = \true, \etc)$\\
\hdashline
    \tab{+}{1}$\sum_{p,p_x{:}P,U{:}\mathcal{P}(P)} (d_p \approx \textit{SAF}_U)$\\
    \tab{$\to$}{2}$\textit{store}(\textit{Forbidden}(p_x), \true, p) \cdot \textit{BF}(\langle d_p = \textit{SAF}_{p_x,U}, \textit{forbidden}[p_x\mapsto \true], \etc)$\\
    \tab{+}{1}$\sum_{p,p_x{:}P,U{:}\mathcal{P}(P)} (d_p \approx \textit{SAF}_{p_x,U} \wedge \textit{busy}(p_x))$\\
    \tab{$\to$}{2}$\textit{load}(\textit{Busy}(p_x), \true, p) \cdot \textit{BF}(\langle d_p = \textit{SAF}_{p_x,U}^{\textit{undo}},\etc)$\\
    \tab{+}{1}$\sum_{p,p_x{:}P,U{:}\mathcal{P}(P)} (d_p \approx \textit{SAF}_{p_x,U} \wedge P \setminus \{p_x\} \approx U \wedge \neg \textit{busy}(p_x))$\\
    \tab{$\to$}{2}$\text{load}(\textit{Busy}(p_x), \false, p) \cdot \textit{BF}(\langle d_p = \textit{LOS}_2,\etc)$\\
    \tab{+}{1}$\sum_{p,p_x{:}P,U{:}\mathcal{P}(P)} (d_p \approx \textit{SAF}_{p_x,U} \wedge P \setminus \{p_x\} \not\approx U \wedge \neg\textit{busy}(p_x))$\\
    \tab{$\to$}{2}$\textit{load}(\textit{Busy}(p_x), \false, p) \cdot \textit{BF} ( \langle d_p = \textit{SAF}_{U \cup \{p_x\}}, \etc)$\\
    \tab{+}{1}$\sum_{p,p_x{:}P,U{:}\mathcal{P}(P)} (d_p \approx \textit{SAF}_{p_x,U})$\\
    \tab{$\to$}{2}$\textit{store}(\textit{Forbidden}(p_x), \false, p) \cdot \textit{BF}(\langle d_p{=}\textit{SAF}_{U\setminus\{p_x\}}, \textit{forbidden}[p_x\mapsto \false], \etc)$\\
    \tab{+}{1}$\sum_{p,p_x{:}P,U{:}\mathcal{P}(P)} (d_p \approx \textit{SAF}^{\textit{undo}}_{p_x,U})$\\
    \tab{$\to$}{2}$\textit{store}(\textit{Forbidden}(p_x), \false, p) \cdot \textit{BF}(\langle d_p{=}\textit{SAF}_{U\setminus\{p_x\}}, \textit{forbidden}[p_x\mapsto \false], \etc)$\\
\hdashline
    \tab{+}{1}$\sump (d_p \approx \textit{LOS}_2) \to \textit{internal} \cdot \textit{BF}(\langle d_p = \textit{LOS}_1, \etc)$\\
    \tab{+}{1}$\sump (d_p \approx \textit{LOS}_1) \to \texttt{enter\_exclusive\_return}(p) \cdot \textit{BF}(\langle d_p = \textit{Exclusive},\etc)$\\
\hdashline
    \tab{+}{1}$\sump (d_p \approx \textit{Exclusive}) \to \texttt{leave\_exclusive\_call}(p) \cdot \textit{BF}(\langle d_p = \textit{LE}_P, \etc)$\\
    \hdashline
    \tab{+}{1}$\sum_{p,p_x{:}P,U{:}\mathcal{P}(P)} (d_p \approx \textit{LE}_U \wedge U \approx \{p_x\})$\\
    \tab{$\to$}{2}$\text{store}(\textit{Forbidden}(p_x), \false, p) \cdot \textit{BF}(\langle d_p = \textit{OE}_1, \textit{forbidden}[p_x\mapsto\false],\etc)$\\
    \tab{+}{1}$\sum_{p,p_x{:}P,U{:}\mathcal{P}(P)} (d_p \approx \textit{LE}_U \wedge U \not\approx \{p_x\})$\\
    \tab{$\to$}{2}$\textit{store}(\textit{Forbidden}(p_x), \false, p) \cdot \textit{BF}(\langle d_p = \textit{LE}_{U\setminus\{p_x\}}, \textit{forbidden}[p_x\mapsto\false],\etc)$\\
    \tab{+}{1}$\sum_{p,p_x{:}P,U{:}\mathcal{P}(P)} (d_p \approx \textit{LE}_U)$\\
    \tab{$\to$}{2}$\textit{store}(\textit{Forbidden}(p_x), \true, p) \cdot \textit{BF}(\langle d_p = \textit{LE}_{U\cup\{p_x\}}, \textit{forbidden}[p_x\mapsto\false], \etc)$\\
\hdashline
    \tab{+}{1}$\sump (d_p \approx \textit{OE}_2) \to \textit{unlock}(p) \cdot \textit{BF}(\langle d_p = \textit{OE}_1, \textit{mtx} = \false, \etc)$\\
    \tab{+}{1}$\sump (d_p \approx \textit{OE}_1) \to \texttt{leave\_exclusive\_return}(p) \cdot \textit{BF}(\langle d_p = \textit{Free}, \etc)$ \\ \hline
    \end{tabular}
    \caption{Linear process equation of the busy-forbidden implementation.}
    \label{spec:busy_forbidden_impl}
\end{table}

\section{Linear process equations of the busy-forbidden protocol}
Table \ref{spec:busy_forbidden_spec} contains the linear process equation of the externally visible behaviour of the busy-forbidden protocol.

\section{Proofs of invariants and reachable states}
In this section of the appendix, we give the explicit proofs of the invariants shown in Section \ref{sec:models_of_busyforbidden}.
We also provide the proof of Theorem \ref{thm:aux_equiv}, in which we show that:
Given a state $d:D$ in which a thread is present in the \textit{SAF} node and such that the focus condition holds and the transition from \textit{SAF} to \textit{LOE} is enabled in the specification. There must be some finite path of $int$ actions from this state to a state $d_{\inta}$ in which the aforementioned action to the \textit{LOE} node is enabled.

\begin{invariant} 
The following invariant holds in the initial state and all subsequent states of the implementation: Given any state $d{:}D$ as per Definition \ref{def:states_impl},
$$\exists p {:} P.~ d_p \in M \Leftrightarrow \textit{mtx} \text{, and }\forall p_x, p_y {:} P .~ d_{p_x},d_{p_y} \in M \Rightarrow p_x = p_y \text{ where }$$
$$ M =\textit{SAF} \cup \textit{LOS} \cup \textit{Exclusive} \cup \textit{LE} \cup \{ \textit{OE}_1\}$$
\end{invariant}
\textit{Proof.} This invariant holds in the initial state as all substates are initially set to \textit{Free} and \textit{mtx} is set to $\false$.

Consider some state $d{:}D$ in which the invariant holds and no thread is present in the given set of substates $M$.
A thread can only enter this set through the $\textit{lock}(p)$ action in \textit{EE}.
This is also the only action that results in \textit{mtx} being set to $\true$, thus the invariant will also hold in each subsequent state of $d$ given that no thread is present in $M$.

Now consider some state $d\textit{:}D$ in which the invariant holds and a (single) thread is present in the given set of substates $\textit{MTX}$.
A thread con only leave this set through the $\textit{unlock}(p)$ action in $\textit{OE}_1$.
This is also the only action that results in \textit{mtx} being set to $\false$, thus the invariant will hold in all subsequent states of $d$. \qed

\begin{invariant} 
The following invariant holds in the initial state and all subsequent states of the implementation: Given any state $d{:}D$ as per Definition \ref{def:states_impl},
$$\forall p{:}P. d_p \in B \Leftrightarrow \textit{busy}(p) \text{ where } B = \textit{LOE} \cup \textit{Shared} \cup \{ \textit{ES}_2,\textit{ES}_3, \textit{LS}_1\} \text{.}$$
\end{invariant}
\textit{Proof.} This invariant holds in the initial state as all substates are initially set to \textit{Free} and \textit{busy} initially maps each thread $p$ to $\false$.

Consider some state $d{:}D$ in which the invariant holds.
A thread $p$ can only enter the set of substates $B$ through the $\textit{store}(\textit{Busy}(p), \true, p)$ action in $\textit{ES}_1$.
This is the only action that results in $\textit{busy}$ mapping $p$ to $\true$.
Thus if there is no thread present in the set of substates $B$, the invariant will also hold after each enabled action.
Similarly, a thread $p$ can only leave the set of substates $B$ through the $\textit{store}(\textit{Busy}(p), \false, p)$ action in either $\textit{ES}_3$ or $\textit{LS}_1$, which are the only actions that result in $\textit{busy}$ mapping $p$ to $\false$.
Thus if the invariant holds in some state $d{:}D$, it also holds in all subsequent states. \qed

\begin{invariant} 
The following invariant holds in the initial state and all subsequent states of the implementation: Given any state $d{:}D$ as per Definition \ref{def:states_impl},
$$\forall p{:}P. \textit{forbidden}(p) \Longleftrightarrow \exists q{:}P. d_q \in F \text{,}$$
where $F = \textit{LOS} \cup \textit{Exclusive} \cup \{\textit{LE}_U | U \subset P \wedge p\in U\} \cup \{\textit{SAF}_U | U \subset P \wedge p \in U\} \cup \{\textit{SAF}_{p,U} | U \subset P \} \cup \{\textit{SAF}_{p,U}^{\textit{undo}} | U \subset P\}$.
\end{invariant}
\textit{Proof.} The invariant holds in the initial state as all substates are initially set to \textit{Free} and \textit{forbidden} initially maps each thread $p$ to $\false$.

Consider some state $d{:}D$ in which the invariant holds.
The set of substates $F$ is a subset of $M$, given in Invariant \ref{inv:exclusive_exclusive}, and thus there can be at most $1$ thread present in $F$.
The only action that results in \textit{forbidden} mapping $p$ to $\true$ is the $\textit{store}(\textit{Forbidden}(p), \true, q)$ action that is enabled in the states $\textit{SAF}_U$ and $\textit{LE}_U$.
This is also the only action that results in some thread $q$ entering $F$.
Thus if there is no thread present in $F$, the invariant will also hold after each enabled action.
Similarly, a thread $q$ can only leave $F$ through the $\textit{store}(\textit{Forbidden}(p), \false, q)$ action in $\textit{SAF}_{p,U}$, $\textit{SAF}_{p, U}^{\textit{undo}}$, and $\textit{LE}_U$, which are the only actions that result in $\textit{forbidden}$ mapping $p$ to $\false$.
Thus if the invariant holds in $d$, it also holds in all subsequent states. \qed

\setcounter{theorem}{1}
\begin{theorem} 
    \normalfont{Given some state $d{:}D$, some thread $p_{\textit{SAF}}{:}P$, and some data configuration $e_\tau{:}E_\tau$ such that $\textit{FC}(d)$ and $c'_\tau(h(d), e_\tau)$ hold, $h(d)(p_{\textit{SAF}} = \textit{SAF}$ and $g'_\tau(h(d), e_\tau) = \textit{LOE}$.
    There must be some state $d_{\inta}{:}D$ such that $d {\xrightarrow{\scriptsize{\inta}}}\!^* d_{\inta}$ and $c_\tau(d_{\inta}, e_\tau)$ hold and $h(g_\tau(d_{\inta}, e_\tau))(p_{\textit{SAF}}) = \textit{LOE}$.}
\end{theorem}
\textit{Proof.} We define the state
    $\textit{aux}_U(d) = \langle d_1', \hdots, d_N', \textit{busy}', \textit{forbidden}', \textit{mutex}'\rangle$, for any given subset of tread $U \subset P$, such that for all $p {:} P$ we have:
    \begin{align*}
        d_p' = \textit{ES}_1 & \text{ iff } d_p \in \textit{ES} \wedge d_p \in U \\
        d_p' = \textit{ES}_2 & \text{ iff } d_p \in \textit{ES} \wedge d_p \not\in U \\
        d_p' = \textit{SAF}_U & \text{ iff } d_p \in \textit{SAF} \\
        d_p' = d_p & \text{ otherwise}
    \end{align*}
The values of $\textit{busy}'$, $\textit{forbidden}'$, and $\textit{mutex}'$ can be inferred from $d_{p_1}', \hdots, d_{p_N}'$ as per the invariants.

We show that $d {\xrightarrow{\scriptsize{\inta}}}\!^* \textit{aux}_U(d)$ holds for any given $U \subset P$ using induction over $U$. 
For the base-case, $U = \emptyset$, we have $d = \textit{aux}_\emptyset$.
Now take any $U \subset P$ such that $d =_{\inta} \textit{aux}_U(d)$ and any $p{:}P$ such that $p \not\in U$ and $U\cup\{p\} \subset P$. 
We show that $\textit{aux}_U(d) =_{\inta} \textit{aux}_{U\cup\{p\}}(d)$.
Take $p_{\textit{SAF}} {:} P$ such that $d_{p_{\textit{SAF}}} = \textit{SAF}_U$. 
If $d_p \in \textit{ES}$, then the sequence of actions $\textit{store}(\textit{Forbidden}(p), \true, p_{\textit{SAF}}) \cdot$ $ \textit{load}(\textit{Forbidden}(p), \true, p) \cdot $ $\textit{store}(\textit{Busy}(p), \false, p) \cdot $ $ \textit{load}(\textit{Busy}(p), \false, p_{\textit{SAF}})$ \\takes us from $\textit{aux}_U(d)$ to $\textit{aux}_{U\cup\{p\}}(d)$. 
If $p \not\in \textit{ES}$, then the sequence of actions $\textit{store}($ $\textit{Forbidden}(p), $ $\true,$ $p_{\textit{SAF}}) \cdot \textit{load}(\textit{Busy}(p),$ $ \false, p_{\textit{SAF}})$ takes us from $\textit{aux}_U(d)$ to $\textit{aux}_{U\cup\{p\}}(d)$. 
As the relation ${\xrightarrow{\scriptsize{\inta}}}\!^*$ is transitive, it follows that $d {\xrightarrow{\scriptsize{\inta}}}\!^* \textit{aux}_U(d)$ for any given $U \subset P$.

We have thus shown that if we take $U = P \setminus \{p_{\textit{SAF}}\}$ such that $p_{\textit{SAF}} \in \textit{SAF}$, we have $d {\xrightarrow{\scriptsize{\inta}}}\!^* \textit{aux}_U(d)$, from which we can take a single \inta action, i.e.\ $\textit{store}(\textit{Forbidden}(p_{\textit{SAF}}),$ $ \true, p_{\textit{SAF}})$, after which the transition to the \textit{LOE} node will be enabled. \qed

\section{Detailed equivalence proof}
We give a more detailed version of the equivalence proof shown in the main paper.
\begin{theorem} 
    \normalfont{The LPE of the implementation given in Table \ref{spec:busy_forbidden_impl} and the LPE of the specification given in Table \ref{spec:busy_forbidden_spec} are divergence-preserving branching bisimilar.}
\end{theorem}
\noindent \textit{Proof.} We show that all ten requirements given in Theorem \ref{thm:new_cnf} hold using Invariants \ref{inv:exclusive_exclusive}, \ref{inv:busy_in_shared}, and \ref{inv:forbidden}, and the state mapping, focus condition, ordering and cone labeling, given in Definitions \ref{def:statemapping}, \ref{def:focuscondition}, \ref{def:order}, and \ref{def:cone_labeling}, respectively.
And thus the implementation and specification are divergence-preserving branching bisimilar.

Requirement \ref{req:bbsim_cones_converge_to_fc} holds since for each substate that does not satisfy the sub-focus condition, at least one internal action with a smaller endpoint, is always enabled.
We can see that none of the \inta actions in the implementation, i.e.\ \textit{italicized} actions in Table \ref{spec:busy_forbidden_impl}, have an endpoint that is in a different cone then their beginpoint.
Thus, Requirement \ref{req:bbsim_closed_under_internal_actions} also holds.

Both the implementation and specification contain exactly three externally visible actions that are not always enabled.
We show that if the action in the implementation is enabled, the action in the specification is also enabled, thus showing that Requirement \ref{req:bbsim_act_in_cone_to_spec} holds.
Simultaneously, we show that if an action in the specification is enabled, the same action is also enabled in the corresponding focus point in the implementation, thus showing that Requirement \ref{req:bbsim_focus_points_mimic_specification_states} holds.

The first action is the $\text{load}(\textit{Forbidden}(p),\false, p)$ action in $\textit{ES}_2$ and the $\tau$ transition from the \textit{ES} to the \textit{LOE} node in the specification.
The \textit{load} action is only enabled when $\textit{forbidden}(p)$ is $\false$, and the $\tau$ transition in the specification is only enabled if there are no threads in \textit{LOS} or \textit{Exclusive} node.
As per Invariant \ref{inv:forbidden}, $\textit{forbidden}(p)$ is $\true$ iff there is a thread present in one of these nodes, and thus when the transition is not enabled in the specification it is also not enabled in the implementation.
As per the same invariant, if the action is enabled in the specification it is also enabled in all corresponding focus points. \\
\indent The second action is the $\text{lock}(p)$ action in \textit{EE} and the $\tau$ transition in the \textit{EE} node in the specification.
The \textit{lock} action is only enabled when \textit{mtx} is $\false$, and the $\tau$ transition in the specification is only enabled if there is no thread in the \textit{SAF}, \textit{LOS}, \textit{Exclusive}, and \textit{LE} node.
As per Invariant \ref{inv:exclusive_exclusive}, if \textit{mtx} is $\false$ iff there is no thread in any of the states belonging to the aforementioned nodes or in the $\textit{OE}_1$ substate.
Thus, if the action is enabled in the implementation it is also enabled in the specification.
Because the focus condition does not hold in $\textit{OE}_1$, we have that if the action is enabled in the specification, it is also enabled in the corresponding focus point in the implementation.\\
\indent The third action is the $\text{load}(\textit{Busy}(p_x),\false, p)$ action in $\textit{SAF}_{p_x,U}$ and the $\tau$ transition from the \textit{SAF} to the \textit{LOS} node in the specification.
The \textit{load} action is only enabled when $\textit{Busy}(p)$ is $\false$ and the $\tau$ transition is only enabled if there is no thread in the \textit{LOE} and \textit{Shared} nodes.
As per Invariant \ref{inv:busy_in_shared}, if $\textit{busy}(p)$ is $\false$ then the \textit{LOE} and \textit{Shared} node are empty and thus, if the action is enabled in the implementation, it is also enabled in the specification.
As per the same invariant, the only focus points in which the action would not be enabled while it would be in the corresponding specifications state, are the ones in which a thread is in the \textit{SAF} node, i.e.\ some thread $p{:}P$ has the substate $\textit{SAF}_\emptyset$.
In these cases, as per Theorem \ref{thm:aux_equiv},there must be some finite path of \inta actions to some state $d_{\inta}$ in which this action is enabled.
The only visible actions with parameters are the eight \texttt{enter/leave\_shared/exclusive\_call/return} actions.
We can thus quickly see that the parameters of these actions match and Requirement \ref{req:bbsim_arguments_match} holds.\\
\indent Similarly, it should be clear from the linear process equations that the endpoints of all externally visible transitions are also related and thus Requirement \ref{req::bbsim_endpoints_are_related} is also met.
Requirement \ref{req:dpbb_cones_loop} holds, as the cone labeling $p$ labels a cone as divergent exactly when one of the three \textit{improbable} loops would be enabled in the specification. \\
\indent In the corresponding focus points for the \textit{SAF} and \textit{LE} cone, there is always at least one internal action enabled.
In the focus point for the \textit{ES} cone, the $\textit{load}(\textit{Forbidden}(p), \true, p)$ action is enabled iff $\textit{forbidden}(p)$ is $\true$.
As per Invariant \ref{inv:forbidden}, the only focus points in which $\textit{Forbidden}(p)$ is $\true$ are the ones in which the \textit{LOS} or \textit{Exclusive} node are occupied.
In all other focus points, there are no further internal actions enabled.
Thus Requirement \ref{req:dpbb_focus_points_diverge} holds. \\
\indent If a cone is labelled as non-diverging ($\nabla$), then each thread should be in one of the following nodes:
\textit{Free}, \textit{LOE}, \textit{Shared}, \textit{LS}, \textit{EE}, \textit{LOS}, \textit{Exclusive}, or \textit{OE}, or \textit{ES}, given that there are no threads present in either \textit{LOS} or \textit{Exclusive}.
With the exception of the $\textit{load}(\textit{Forbidden}(p), \true, p)$ action in the \textit{ES} node, all the internal actions within these nodes take us closer to a focus point.
As per Invariant \ref{inv:forbidden}, \textit{forbidden} is $\true$ only if there is a thread present in either the \textit{LOS} or \textit{Exclusive}, \textit{LE}, or \textit{SAF} node, which are known to be empty.
Thus Requirement \ref{req:dpbb_non_divergent_cones_strictly_converge} also holds and the implementation and specification are divergence-preserving branching bisimilar as per Theorem \ref{thm:new_cnf}. \qed

\end{document}